\documentclass[preprint]{aastex61}

\def\ME{M_\oplus}

\UseRawInputEncoding

\shorttitle{Physical conditions of gas components in debris disks of 49~Ceti and HD~21997}
\shortauthors{Higuchi et al.}

\begin{document}
\title{Physical conditions of gas components in debris disks of 49~Ceti and HD~21997}

\correspondingauthor{Aya E. Higuchi}
\email{aya.higuchi@nao.ac.jp}

\author[0000-0002-9221-2910]{Aya E. Higuchi}
\affiliation{National Astronomical Observatory of Japan, 2-21-1 Osawa, Mitaka, Tokyo 181-8588, Japan}

\author{\'Agnes K\'osp\'al}
\affiliation{Konkoly Observatory, Research Centre for Astronomy and Earth Sciences, Konkoly-Thege Mikl\'os \'ut 15-17, 1121 Budapest, Hungary}
\affiliation{Max Planck Institute for Astronomy, K\"onigstuhl 17, D-69117 Heidelberg, Germany}
\affiliation{ELTE E\"otv\"os Lor\'and University, Institute of Physics, P\'azm\'any P\'eter s\'et\'any 1/A, 1117 Budapest, Hungary}

\author{{Attila Mo{\'o}r}}
\affiliation{Konkoly Observatory, Research Centre for Astronomy and Earth Sciences, Konkoly-Thege Mikl\'os \'ut 15-17, 1121 Budapest, Hungary}
\affiliation{ELTE E\"otv\"os Lor\'and University, Institute of Physics, P\'azm\'any P\'eter s\'et\'any 1/A, 1117 Budapest, Hungary}

\author{Hideko Nomura}
\affil{National Astronomical Observatory of Japan, 2-21-1 Osawa, Mitaka, Tokyo 181-8588, Japan}
\affiliation{The Graduate University for Advanced Studies (SOKENDAI), 2-21-1 Osawa, Mitaka, Tokyo 181-0015, Japan}

\author{Satoshi Yamamoto}
\affiliation{Department of Physics, The University of Tokyo, Hongo, Bunkyo-ku, Tokyo 113-0033, Japan}

\begin{abstract}

Characterization of gas component in debris disks is of fundamental importance for understanding its origin.
Toward this goal, we have conducted non-LTE (local thermodynamic equilibrium) analyses of the rotational spectral lines of CO including
those of rare isotopologues ($^{13}$CO and C$^{18}$O) observed toward the gaseous debris disks of 49~Ceti and HD~21997
with the Atacama Large Millimeter/submillimeter Array (ALMA) and Atacama Compact Array (ACA).
The analyses have been carried out for a wide range of the H$_{2}$ density, and the observed
line intensities are found to be reproduced, as far as the H$_{2}$ density is higher than 10$^{3}$~cm$^{-3}$.
The CO column density and the gas temperature are evaluated to be (1.8--5.9)$\times$10$^{17}$~cm$^{-2}$ and 8--11~K for 49~Ceti and 
(2.6--15)$\times$10$^{17}$~cm$^{-2}$ and 8--12~K for HD~21997, respectively, where the H$_{2}$ collision is assumed for the rotational excitation of CO.
The results do not change significantly even if electron collision is considered.
Thus, CO molecules can be excited under environments of no H$_{2}$ or a small number of H$_{2}$ molecules,
even where the collision with CO, C, O, and C$^{+}$ would make an important contribution for the CO excitation in addition to H$_{2}$.
Meanwhile, our result does not rule out the case of abundant H$_{2}$ molecules.
The low gas temperature observed in the debris disks is discussed in terms of inefficient heating by interstellar and stellar UV radiation.


\end{abstract}
\keywords{Circumstellar matter: Debris disks: Observational astronomy: Submillimeter astronomy}

\section{Introduction}

Gas component of debris disks has attracted many astronomers and planetary scientists in relation to evolutionary 
process from protoplanetary disks to planetary system.
It has been detected in optical absorption lines of some atoms \cite[e.g.,][]{sle75, rob00, rob06}, far-infrared emission lines of 
[O\,{\sc i}] and [C\,{\sc ii}] \cite[e.g.,][]{don13,rob13,rob14,riv12,riv14,cat14,bra16}, and millimeter/submillimeter emission 
lines of CO \citep{zuc95,den05,moo11,moo15}.
With the advent of the Atacama Large Millimeter/submillimeter Array (ALMA), 
the CO emissions have extensively been studied in many debris disks during recent years 
\cite[e.g.,][]{kos13, den14, lie16, hug17, moo17, mat17, hug18, moo19}.
Not only $^{12}$CO but also its isotopic species ($^{13}$CO, and C$^{18}$O) have successfully been observed in a few sources, 
e.g., HD~21997, HD~121191, HD~121617, HD~131488, 49~Ceti, and HD~32297 \citep{kos13, moo17, moo19}.
In addition, the [C\,{\sc i}] emission has been detected around 49~Ceti \citep{hig17,hig19a}, 
$\beta$ Pictoris \citep{hig17,cat18}, HD~131835 \citep{kra19}, and HD~32297 \citep{cat20}.
Very recently, detection of [$^{13}$C\,{\sc i}] has been reported in 49~Ceti \citep{hig19b}.
CO emissions in debris disks have often been assumed to be optically thin, 
because its intensities from several gaseous debris disks are faint even with ALMA observations.
However, detection of $^{13}$CO and C$^{18}$O in a few sources 
\citep{kos13, moo17, moo19} clearly indicates that the $^{12}$CO emission is no longer regarded as optically thin in these gas disks.
This situation is also true for the [C\,{\sc i}] emission \citep{hig19a, hig19b}.

Based on the observed peak intensities of optically thick $^{12}$CO and $^{13}$CO lines, 
\cite{kos13} suggested excitation temperatures as low as 6--9~K in the disk of HD~21997. 
Assuming LTE (local thermodynamic equilibrium) they derived a total CO mass of about (4--8)$\times$10$^{-2}$$\ME$.
Since the rotational temperature is much lower than the dust temperature reported so far \citep{moo13, hol17}, 
such a low excitation temperature raised an important issue on the physical condition
and the amount of the gas component in debris disks.
Later, similar trends of the low excitation temperature were reported for several debris disks \citep{fla16, moo17, di20}.
Since the temperature is derived under the LTE assumption, 
it was thought that the non-LTE effect may affect the excitation temperature \citep{mat15}. 
For 49~Ceti, \cite{hug17} analyzed the $^{12}$CO($J$=3--2) data observed with ALMA and the $^{12}$CO($J$=2--1) data 
observed with the Submillimeter Array (SMA) by using the non-LTE code along with the disk model, and derived the temperature structure of the form as:
\begin{equation}
\label{eq1}
T(R) = T_{100}\left(\frac{R}{100~\rm{au}}\right)^{-0.5},
\end{equation}
where $R$ is the radial distance from the central star and $T_{100}$ is the gas temperature at $R$=100~au.
$T_{100}$ is evaluated to be 40~K and 14~K for the H$_{2}$/CO ratio of 10$^{4}$ and 1, respectively.
They also reported that the CO lines in 49~Ceti are almost thermalized.
Similar analyses have been reported for the other debris disks \citep{mat15,hal19}.
However, these studies make use of only the $^{12}$CO data.
For further constraint of the gas kinetic temperature, it is essential to involve the $^{13}$CO and C$^{18}$O data in the analysis.

Moreover, the CO abundance relative to H$_{2}$ is still uncertain, although it is deeply related to the origin of the gas
(i.e., a primordial gas inherited from the protoplanetary disk and/or a secondary gas outgassing from dust grains and icy solids).
\cite{hig17} pointed out, by using a simple chemical model, that the C/CO column density ratio (hereafter C/CO ratio) is 
sensitive to the H$_{2}$ number density.
To make full use of this method, we need to derive the C/CO ratio accurately by eliminating the effect of the opacity
of the CO and [C\,{\sc i}] emission.

In order to constrain the gas kinetic temperature and the CO column density accurately, 
use of as many molecular lines as possible is essential.
It is particularly important to include optically thin lines such as rare isotopologue lines.
To demonstrate this, we here focus on 49~Ceti and HD~21997, which are nearby 
(57.0$\pm$0.3~pc ;\cite{gai18,bai18} for 49~Ceti, 69.5$\pm$0.2~pc ;\cite{bai18} for HD~21997) 
famous A-type stars (A1 for 49~Ceti, A3 for HD~21997) at the ages of 40 -- 50~Myr for 49~Ceti \citep{zuc12, zuc19} 
and 30 -- 45~Myr for HD~21997 \citep{moo06,tor08,bel15}.
There are several high-sensitivity ALMA archival data for these two sources listed in Table \ref{tb1}.
In this study, we have reanalyzed these archival data sets of 49~Ceti and HD~21997 to derive the column 
density and the gas temperature for observational understandings of the gas properties in these sources apart from the disk modeling.

\section{Data Reduction}

\subsection{49~Ceti}

We used the ALMA archival data sets of 49~Ceti obtained by using the Band 6 and Band 7 receivers (Table \ref{tb1}).
For a fair comparison among the observations with different spatial resolution, the synthesized beam size
for the 12~m array data is adjusted to that for the ACA data ($\sim$6$\arcsec$) by using the CASA task $\it{imsmooth}$.
The synthesized beam sizes are summarized in Table \ref{tb1}.
The angular size of 6$\arcsec$ corresponds to 340~au.

We used version 4.7.2 and 5.3.0 of the Common Astronomy Software Applications package (CASA) \citep{mcm07} 
for calibration and imaging, respectively.
The CASA task $\it{tclean}$ was employed to Fourier-transform the visibility data and to deconvolve the dirty images 
at a velocity interval of 2~km~s$^{-1}$.
Briggs weighting of +0.5 was applied to all line images for the best compromise between resolution and sensitivity.
Continuum subtraction was performed for all the data by using the CASA task $\it{uvcontsub}$.

\subsection{HD~21997}

For HD~21997, we used the ALMA archival data (reduced cube data) reported by \cite{kos13} (Table \ref{tb1}).
The synthesized beam is adjusted to be 4\farcs0 $\times$ 4\farcs0 with P.A.= 0.0$^{\circ}$ for all the CO lines 
to cover the most of the CO distribution.
For this source, the angular size of 4$\arcsec$ corresponds to 280~au.

\section{Results and Discussions}

\subsection{Overall data}

\subsubsection{49~Ceti}

\begin{figure}
\rotate
\epsscale{1}
\plotone{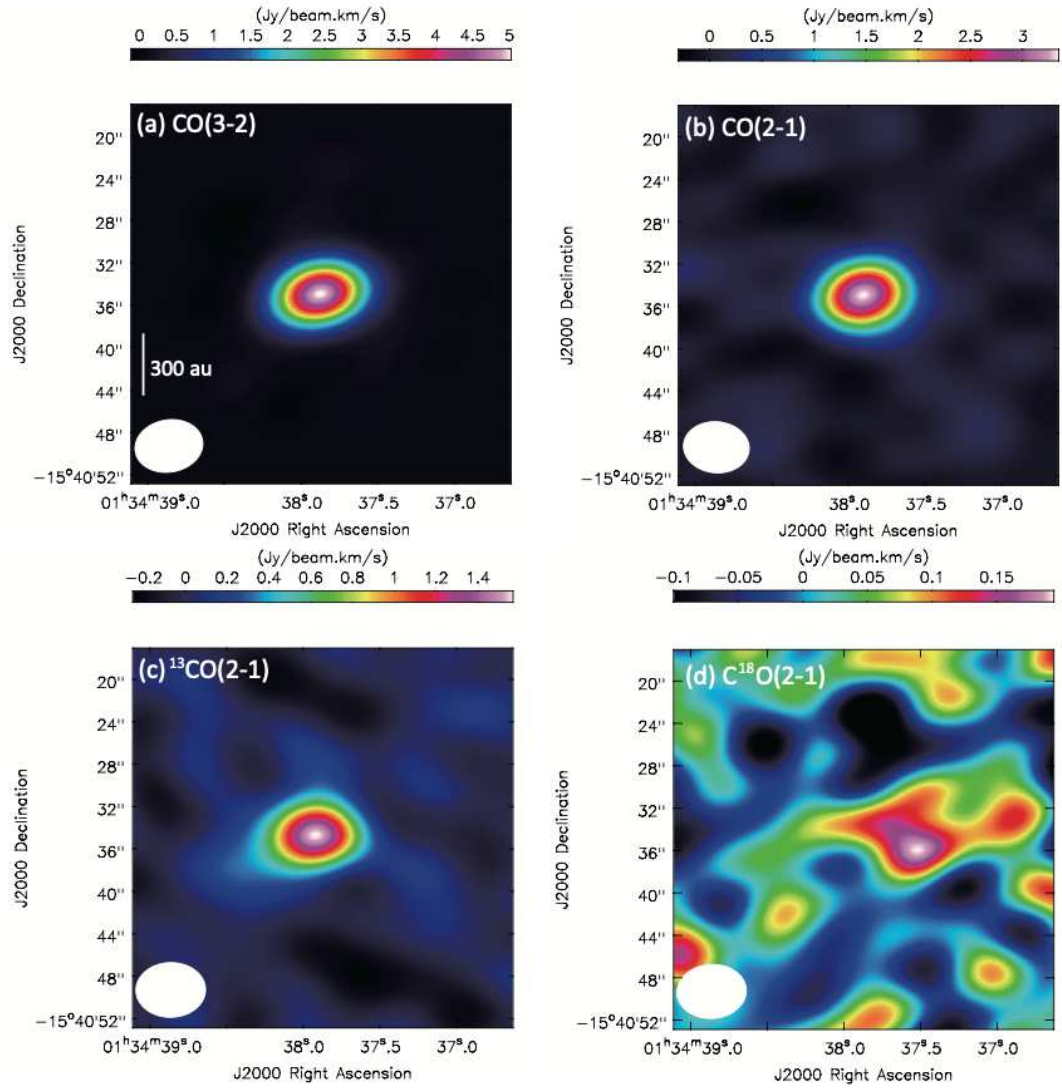}
\caption{(a) Integrated intensity map of the $^{12}$CO(3--2) emission of 49 Ceti obtained by ALMA 12~m observations 
\citep{hug17}. Spatial resolution was smoothed to match that of the ACA observations.
(b): Integrated intensity map of the $^{12}$CO(2--1) emission obtained by ACA observations \citep{moo19}.
(c): Integrated intensity map of the $^{13}$CO(2--1) emission.
(d): Integrated intensity map of the C$^{18}$O(2--1) emission.}
\label{fig1}
\end{figure}

\begin{figure}
\epsscale{1.1}
\plotone{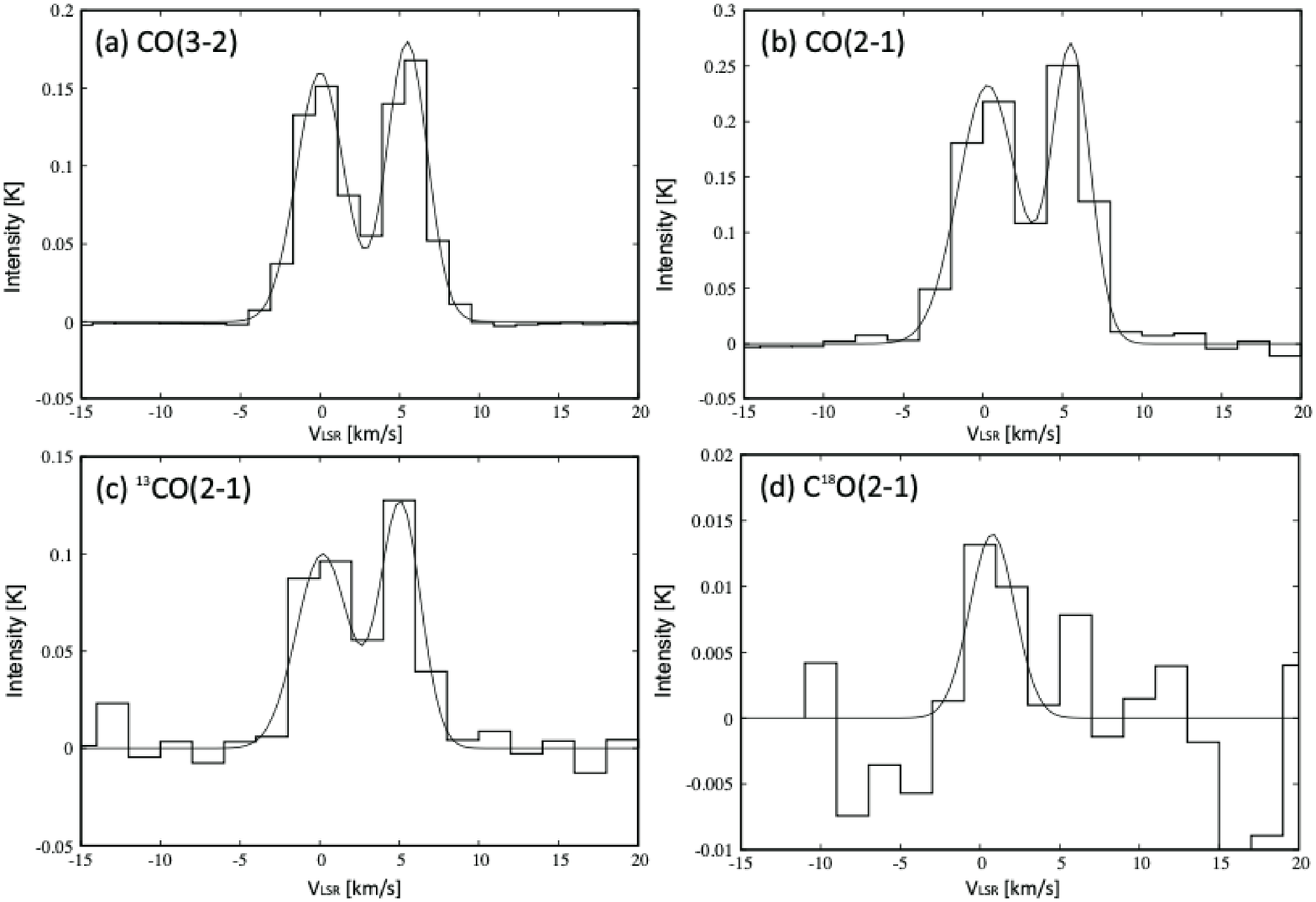}
\caption{
(a) $^{12}$CO(3--2) spectrum of 49~Ceti observed with ALMA 12~m array
\citep{hug17}. The synthesized beam size for the 12~m array data is adjusted to that for the ACA data.
(b) $^{12}$CO(2--1) spectrum of 49~Ceti observed with the ACA.
(c) $^{13}$CO(2--1) spectrum of 49~Ceti. 
(d) C$^{18}$O(2--1) spectrum of 49~Ceti. 
The synthesized beam size for the 12~m array data is adjusted to that for the ACA data.
The solid line indicates the result of double-Gaussian fitting.}
\label{fig2}
\end{figure}

Figures \ref{fig1} (a) - (c) show the integrated intensity maps of the $^{12}$CO(3--2), $^{12}$CO(2--1), 
and $^{13}$CO(2--1) emissions, respectively, at a 6$\arcsec$ resolution, where the velocity range for integration is from $-$6 to 11.5~km~s$^{-1}$.
Figure \ref{fig1} (d) shows the integrated intensity map of the C$^{18}$O(2--1) emission at a 6$\arcsec$ resolution, 
where the velocity range for integration is from $-$4 to 6~km~s$^{-1}$.
Figures \ref{fig2} (a) - (d) show the averaged spectra of the $^{12}$CO(3--2), $^{12}$CO(2--1), 
$^{13}$CO(2--1), and C$^{18}$O(2--1) emissions, respectively.
These spectra are prepared by averaging over the 6$\arcsec$ area in diameter which corresponds to the spatial resolution.
The velocity range for the integrated intensity maps mentioned above is justified by these spectra.
Each spectrum reveals a double-peak profile.
Except for the C$^{18}$O(2--1) data, the intensity of the red-shifted component is brighter 
than that of the blue-shifted component. 
The C$^{18}$O(2--1) line is so faint that the data quality is not as good as the other lines.
Double-Gaussian fitting is performed on all the spectra except for C$^{18}$O(2--1) to derive the line parameters.
For C$^{18}$O(2--1), only the blue shifted component is fitted by the single Gaussian function.
The results are listed in Table \ref{tb2}
\footnote[1]{The conversion equation from Jy~beam$^{-1}$ to K is given as : \\
$\left(\frac{T_{\rm{B}}}{\rm{K}}\right) = 1.222\times10^{6}
\left(\frac{\theta_{\rm{maj}}}{\rm{arcsec}}\right)^{-1}
\left(\frac{\theta_{\rm{min}}}{\rm{arcsec}}\right)^{-1}
\left(\frac{\nu}{\rm{GHz}}\right)^{-2}
\left(\frac{S}{\rm{Jy~beam^{-1}}}\right)$, \\
where $\theta_{\rm{maj}}$ is the major axis, $\theta_{\rm{min}}$ is the minor axis of the synthesized beam,
$\nu$ is the frequency, and $S$ is the flux density.}.

\subsubsection{HD~21997}

\begin{figure}
\rotate
\epsscale{0.8}
\plotone{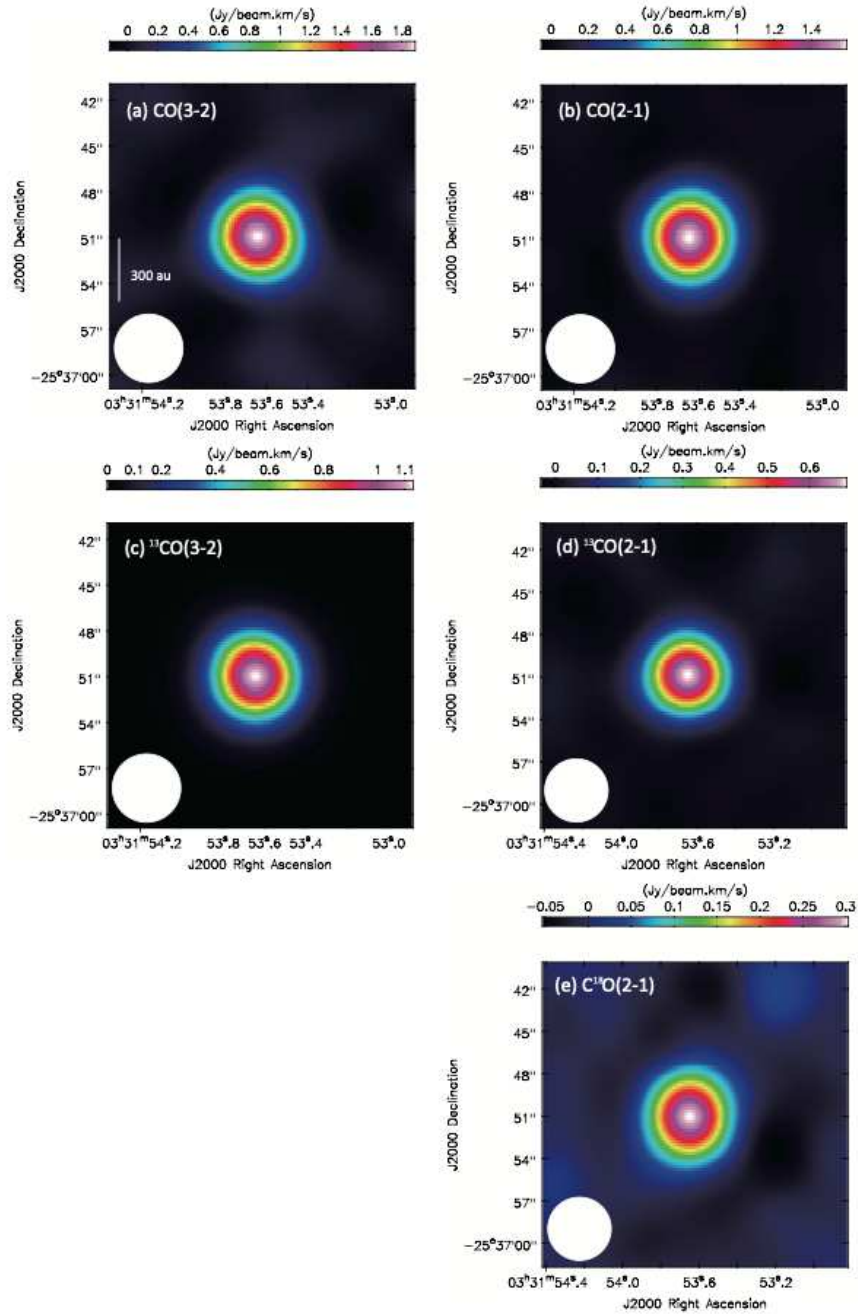}
\caption{
(a) Integrated intensity map of the $^{12}$CO(3--2) emission of HD~21997 obtained by ALMA 12~m observations.
(b): Integrated intensity map of the $^{12}$CO(2--1) emission.
(c): Integrated intensity map of the $^{13}$CO(3--2) emission.
(d): Integrated intensity map of the $^{13}$CO(2--1) emission.
(e): Integrated intensity map of the C$^{18}$O(2--1) emission.}
\label{fig3}
\end{figure}

\begin{figure}
\epsscale{1.1}
\plotone{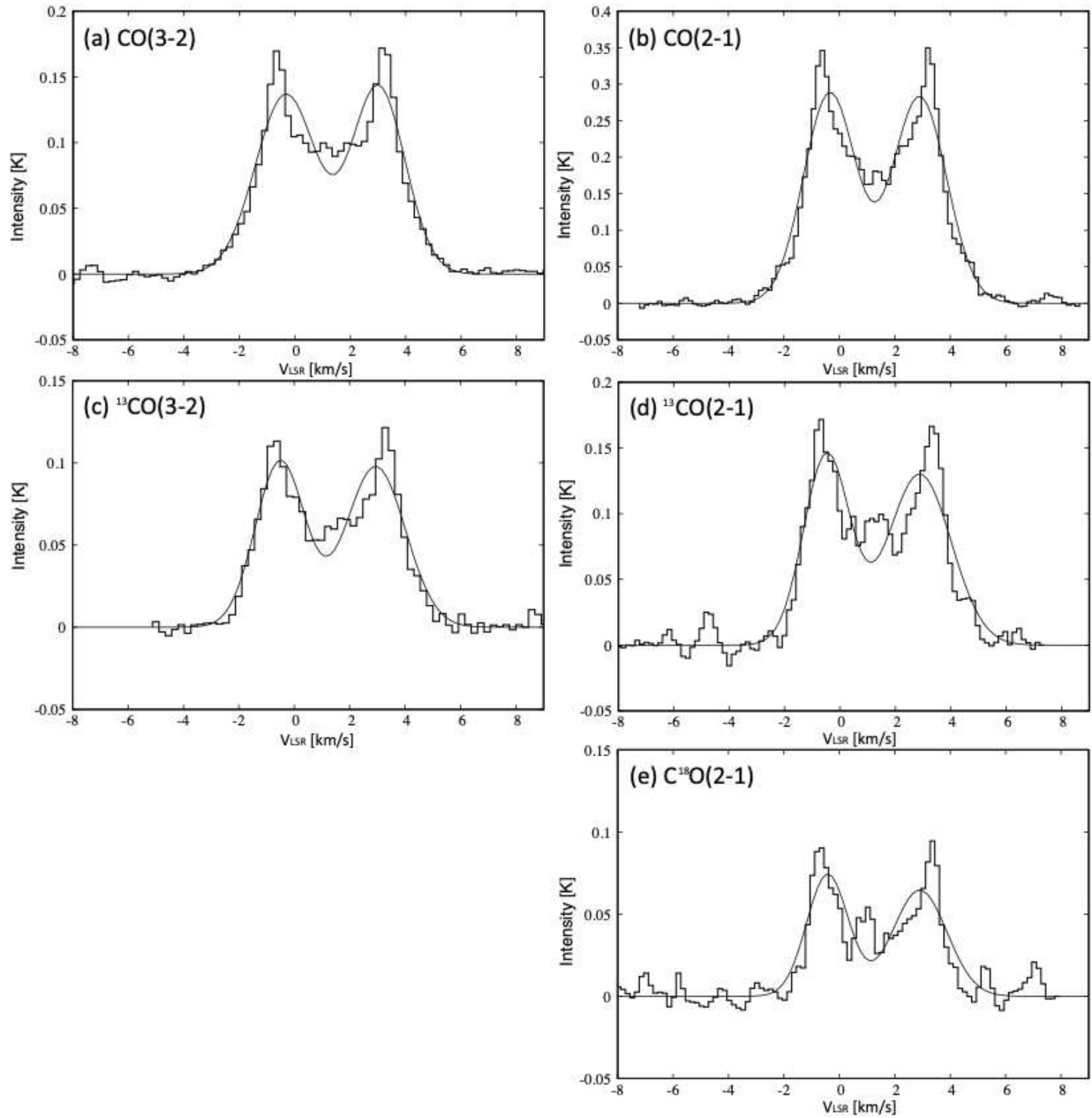}
\caption{
(a) $^{12}$CO(3--2) spectrum of HD~21997 observed with ALMA 12~m array.
(b) $^{12}$CO(2--1) spectrum of HD~21997.
(c) $^{13}$CO(3--2) spectrum of HD~21997.
(d) $^{13}$CO(2--1) spectrum of HD~21997.
(e) C$^{18}$O(2--1) spectrum of HD~21997.
The solid line indicates the result of double-Gaussian fitting.}
\label{fig4}
\end{figure}

Figures \ref{fig3} (a) - (e) show the integrated intensity maps of the $^{12}$CO(3--2), $^{12}$CO(2--1), 
$^{13}$CO(3--2), $^{13}$CO(2--1), and C$^{18}$O(2--1) emissions, respectively, at a 4$\arcsec$ resolution, where the 
velocity range for integration is from $-$4 to 6~km~s$^{-1}$.
Details of the dataset are described elsewhere \citep{kos13}.
Figures \ref{fig4} (a) - (e) show the averaged spectra of the $^{12}$CO(3--2), $^{12}$CO(2--1), $^{13}$CO(3--2), 
$^{13}$CO(2--1), and C$^{18}$O(2--1) emissions, respectively.
These spectra are prepared by averaging over the 4$\arcsec$ area in diameter which corresponds to the spatial resolution.
Since all the spectra show a double-peak emission, double-Gaussian fitting is performed on the spectra to derive the line parameters.
The results are listed in Table \ref{tb3}.
For the $^{13}$CO(3--2), $^{13}$CO(2--1) and C$^{18}$O(2--1) emissions, there seem to be small residuals 
(e.g., Figures \ref{fig4} (b) 0.03~K, (d) 0.02~K and (e) 0.03~K) around the systemic velocity, 
which cannot be reproduced in the double-Gaussian fitting.
Since it is close to the noise level, its contribution to the total spectra is not large.
Thus, we ignore this feature in the following analyses.

\subsection{LTE analysis}

According to the result of double-Gaussian fitting, the peak intensity ratio of $^{13}$CO(2--1)/$^{12}$CO(2--1) 
is calculated to be 0.46$\pm$0.03 and 0.48$\pm$0.04 for 49~Ceti and HD~21997, respectively.
This ratio clearly indicates that the $^{12}$CO(2--1) emission is optically thick, if the $^{12}$C/$^{13}$C ratio of 77 \citep{wil94} is assumed.
A large optical depth of the CO line was already reported by \cite{kos13} for HD~21997.
It is now revealed in 49~Ceti for the first time.

For a more detailed treatment, the rotational temperature, the column density of $^{12}$CO, and the beam filling factor 
are derived by using the following equation, where we take the effect of the optical depth into account:
\begin{equation}
T_{\rm{B}}=f_{d}[B_{\rm{\nu}}(T_{\rm{rot}})-B_{\rm{\nu}}(T_{\rm{CMB}})](1-\exp(-\tau)).
\end{equation}
Here, $T_{\rm{B}}$ is the observed brightness temperature,
$T_{\rm{rot}}$ is the rotational temperature which is equivalent with the excitation temperature in the LTE condition, 
$B_{\rm{\nu}}(T_{\rm{rot}})$ is the Planck function, 
$T_{\rm{CMB}}$ is the temperature of the cosmic background radiation,
and $f_{d}$ is the beam filling factor.
The optical depth, $\tau$, is written as below \cite[e.g.,][]{gol99,sak08}:

\begin{equation}
\tau= \frac{c^{2}}{8\pi\nu^{2}} N_{u} A_{ul} \exp\left(\frac{h\nu}{kT_{\rm{rot}}}\right) 
(1-\exp\left(\frac{h\nu}{kT_{\rm{rot}}}\right)) \sqrt{\frac{4ln2}{\pi}} \frac{1}{\Delta{v}},
\end{equation}
and
\begin{equation}
N_{u}=Ng_{u} \exp\left(\frac{-E_{\rm{u}}}{kT_{\rm{rot}}}\right)/U(T_{\rm{rot}}),
\end{equation}
where $A_{ul}$ is the Einstein A-coefficient of the transition between the states $u$ and $l$, 
$N$ is the column density along the line of sight, $g_{u}$ is the degeneracy of the upper state, 
$E_{u}$ is the upper state energy, $\Delta{v}$ is the velocity width, and $U(T_{\rm{rot}})$ is the partition function.
We employ the isotope ratios of 1/77 and 1/560 for $^{13}$CO and C$^{18}$O, respectively \citep{wil94}.

The spatial distributions and the velocity structures of the $^{12}$CO(3--2), $^{12}$CO(2--1) and $^{13}$CO(2--1) emissions are 
well correlated with one another (see Figures \ref{fig1} -- \ref{fig4}), and hence, the rotational temperature, the column density, 
and the beam filling factor of CO for 49~Ceti and HD~21997 are evaluated by using the least-squares 
analysis on the observed intensities of CO and its isotopic species.
We treat the data for the blue-shifted and red-shifted components separately.
The fitting results (i.e., calculated intensities and residuals in the fit)
are shown in Tables \ref{tb2} and \ref{tb3}, for 49~Ceti and HD~21997, respectively, 
whereas the derived parameters ($T_{\rm{rot}}$, $N$, and $f_{d}$) are summarized in Tables \ref{tb4} and \ref{tb5}, respectively.
The observed peak intensities are well reproduced within the measurement uncertainties for both sources.
The highest correlation among the three parameters in the fit is 0.99 (correlation coefficient between parameters of the least squares analysis) 
between $T_{\rm{rot}}$ and $f_{d}$. 
Nevertheless, these two parameters are separately determined in the fit, and the effect of the correlation is reflected in the quoted errors.

\begin{figure}
\epsscale{1.0}
\plotone{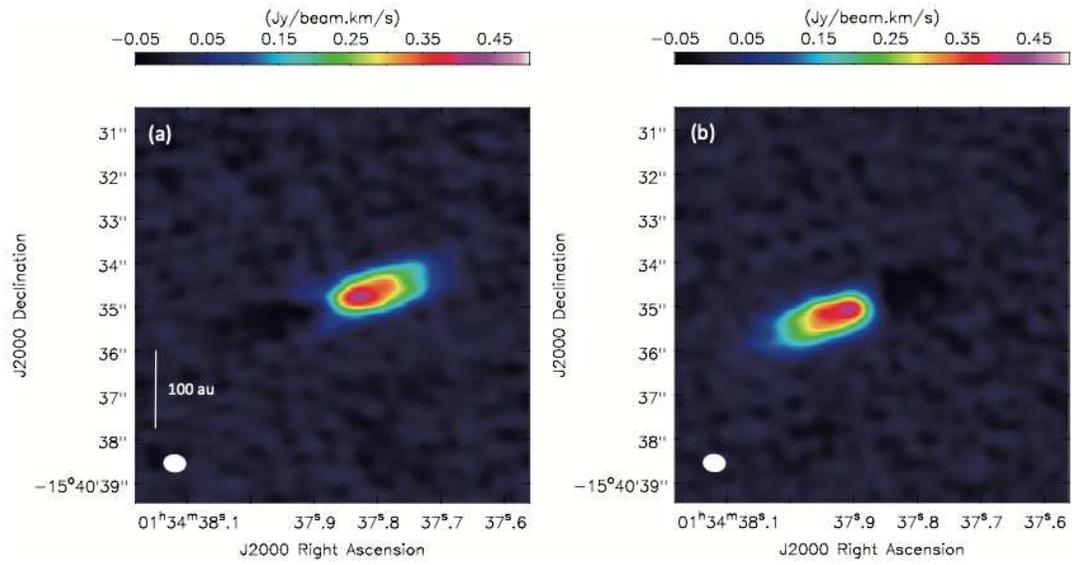}
\caption{Velocity channel maps of the CO(3--2) emissions of (a) the blue-shifted and 
(b) the red-shifted components observed toward 49~Ceti at a high spatial resolution.}
\label{app1}
\end{figure}

For 49~Ceti, the optical depths of the $^{12}$CO(2--1) and $^{12}$CO(3--2) emissions are evaluated to be 38--46 and 18--24, respectively 
(Table~\ref{tb4}).
These $^{12}$CO lines are indeed optically thick.
The excitation temperature is also derived to be $\sim$~9~K from our LTE analysis.
According to the high resolution image (Figure \ref{app1}) reported by \cite{hug17}, 
the FWHM (full width half maximum) size of the emitting region of the CO(3--2) lines is estimated to
be 1\farcs9 $\times$ 0\farcs8 and 1\farcs8 $\times$ 0\farcs8 for the blue-shifted and red-shifted components, respectively,
by using 2D Gaussian fits.
Hence, the beam filling factor is estimated to be 0.04 for both blue-shifted and red-shifted components.
These values are almost consistent with those determined in the fit 
(0.06$\pm$0.01 and 0.08$\pm$0.04 for the blue-shifted and red-shifted components, respectively).

For HD~21997, the optical depths of the $^{12}$CO(2--1) and $^{12}$CO(3--2) 
emissions are 66 -- 76 and 26 -- 36, respectively (Table~\ref{tb5}).
The excitation temperature of gas component in HD~21997 is determined to be 8~K.
Since the emitting region of the CO emission is derived to be 1\farcs1 $\times$ 0\farcs9 and 1\farcs0 $\times$ 0\farcs9 
for the blue-shifted and red-shifted components, respectively, by using 2D Gaussian fits on 
the high resolution channel maps shown in Figure 3 of \cite{kos13}, the beam filling factor is estimated to be 0.06.
This value is consistent with those determined in the fit 
(0.08$\pm$0.06 and 0.07$\pm$0.05 for the blue-shifted and red-shifted components, respectively), 
although the observed beam filling factors suffer from a large error.

For both sources, the beam-averaged excitation temperature (i.e., rotational temperature) is confirmed 
to be lower than 10~K based on the multi-line LTE analysis.
In general, the rotational temperature derived from the LTE analysis does not always correspond to the gas kinetic temperature, 
if the rotational level population is not well thermalized.
To derive the gas kinetic temperature, a non-LTE analysis is performed.

\subsection{Non-LTE analysis}

We employ the non-LTE code prepared by ourselves in the analysis, which is tested 
by using the results reported by \cite{gol74} and also those derived from the RADEX code. 
The collisional cross sections are taken from The Leiden Atomic and Molecular Database (LAMDA)  \citep{sch05}.
In this case, we need to consider the major collision partner for the rotational excitation of CO.
If the gas is mainly primordial (i.e., remnant of the protoplanetary disk), collision with H$_{2}$, 
the way most abundant constituent of such gas material, would be dominant.
On the other hand, if the gas has an almost secondary origin (i.e., releases from icy grains and planetesimals), 
collision with H and O produced by the photodissociation of H$_{2}$O, the C atom, the C$^{+}$ ion, electron, 
and CO itself can be considered.
Collisional rate coefficients for H are lower by an order of magnitude than that for H$_{2}$ \citep{yan13,wal15a}, while those for an 
electron are higher by two orders of magnitude than that for H$_{2}$ \citep{dic77}.
On the other hand, collisional rate coefficients for C, O, CO and C$^{+}$ are not available, but they are roughly assumed to be comparable to those for H$_{2}$.
Based on these considerations, we regard an electron as the major collisional partner for the secondary origin case.
Thus, we conduct the non-LTE analysis for the two distinct cases, where the major collision partner is H$_{2}$ or an electron.
Caveats for this simplification are discussed later.


\subsubsection{The case of H$_{2}$ collision}

First, we conduct the non-LTE analysis based on the LVG (large velocity gradient) model \citep{gol74, van07}, 
assuming the collision with H$_{2}$.
In this analysis, the rotational level populations are determined by the balance between the collision and radiation processes,
from which the line intensities are calculated.
Hence, there are four parameters to be determined:
the H$_{2}$ density, the column density of CO, the gas kinetic temperature, and the beam filling factor.
We have initially tried to optimize these four parameters by using the least-squares fit to reproduce the $^{12}$CO, $^{13}$CO and C$^{18}$O data. 
However, all the four parameters cannot be determined simultaneously in the fit.
Specifically, the H$_{2}$ density is not well constrained.
Hence, we fix the H$_{2}$ density and determine the remaining three parameters.
The range of the H$_{2}$ density assumed in the fit is from 3$\times$10$^{3}$ to 1$\times$10$^{7}$~cm$^{-3}$.
The data for the blue-shifted and red-shifted components are treated separately.
The least-squares fit does not converge for any of the two sources for the H$_{2}$ density below 10$^{3}$~cm$^{-3}$.
Thus, we set the lowest H$_{2}$ density to be 3$\times$10$^{3}$~cm$^{-3}$, for which the fit is successful.
The fitting results (i.e., calculated intensities and residuals in the fit)
are shown in Tables \ref{tb2} and \ref{tb3}, and derived parameters are summarized in Tables \ref{tb6} and \ref{tb7} 
for 49~Ceti and HD~21997, respectively.
The optical depths of the $^{12}$CO, $^{13}$CO, and C$^{18}$O lines and residuals of the fit are also listed.
The fitting is successful, as shown by the residuals.

Here, we assume that the major collisional partner with CO is H$_{2}$ molecules.
However, collision with CO, C, C$^{+}$, and electron(e) would also contribute to the excitation and de-excitation in debris disks.
This is particularly important for the low H$_{2}$ density case.
Since the collisional rate coefficients for the collision with CO/C/C$^{+}$/O are not available,
we cannot distinguish the collision with H$_{2}$ and that of CO/C/C$^{+}$/O in our calculation and just employ the collisional
rate coefficients for the H$_{2}$ collision.
The rate for the CO-H collision is much lower than that of the CO-H$_{2}$ collision \citep{yan13,wal15a}, and thus we have ignored 
the CO-H collision in this study. 
For the contribution of CO, 
the above assumption that the collisional rate coefficients of H$_{2}$ are similar to those of CO could be justified by the 
collisional broadening experiment of the CO line in the laboratory:
broadening of the CO line by H$_{2}$ and self-broadening are similar to each other, indicating that the collisional rate is not very much 
different between H$_{2}$ and CO \citep{dic09}.
In short, we have to recognize the derived H$_{2}$ density as effective one involving the contribution of CO,
C, C$^{+}$ and O: namely, it should approximately be regarded as $n$(H$_{2}$)+$n$(CO)+$n$(C)+$n$(C$^{+}$)+$n$(O).
Hereafter, we denote it as the gas density.
Note that if the abundance of electron is less than 1$\%$ of the above sum, its effect can practically be ignored (Section 3.3.2).

For 49~Ceti, the column density of CO ranges from 5.9$\times$10$^{17}$ to 2.0$\times$10$^{17}$~cm$^{-2}$ for
the blue-shifted component and from 5.4$\times$10$^{17}$ to 1.8$\times$10$^{17}$~cm$^{-2}$ for the red-shifted component (Table \ref{tb6}).
The derived CO column density is higher by 4 orders of magnitude than the upper limit derived from the CO absorption \citep{rob14}.
The column density and optical depth of CO are higher for the lower gas density.
The optical depth is as high as 110 -- 130 for the gas density of 3$\times$10$^{3}$~cm$^{-3}$.
On the other hand, the gas kinetic temperature is between 8 and 11~K and is not very sensitive to the assumed the gas density.
It is similar to the rotational temperature obtained in the LTE analysis.
This means that the level population is almost thermalized in the assumed range of the gas density.
The beam filling factor is almost independent of the gas density and is consistent with that expected 
from the source size and the beam size described in Section 3.2.
Although the maximum correlation in the fit is 0.99 between ${T_{\rm{kin}}}$ and $f_{d}$, 
these parameters are well constrained.

We derive the number density of CO from the column density by assuming the line-of-sight length of the emitting region.
Since 49~Ceti has an almost edge-on configuration, the disk radius of 100~au is employed as the line-of-sight length. 
Then, the averaged CO number density is roughly estimated to be 400~cm$^{-3}$ and 360~cm$^{-3}$ for the 
blue-shifted and red-shifted components, respectively, for the gas density of 3$\times$10$^{3}$~cm$^{-3}$ (Table \ref{tb6}).
Since the number densities of C and C$^{+}$ are expected to be higher than the CO density by about an order of
magnitude \citep{hig17}, the gas density can be interpreted by contributions of CO, C, and C$^{+}$ even without H$_{2}$ molecules in this case.
Note that the lower limit of the [C\,{\sc ii}]  mass is reported to be 2.15$\times$10$^{-4}$~$\ME$ from the Herschel observation \citep{rob13}. 
This lower limit is lower than the CO mass derived above. 
Since the spatial and velocity resolutions of the Herschel observation are much different from our observation, 
it seems difficult to evaluate the contribution of C$^{+}$. If the abundance of C$^{+}$ are low, CO and C would contribute to the excitation.

The number density of CO and H$_{2}$/CO ratio is derived in this way for the other H$_{2}$ density cases, as summarized in Table \ref{tb6}.
If the gas density is lower than 10$^{6}$~cm$^{-3}$, the H$_{2}$/CO ratio is lower than the canonical interstellar value of 10$^{4}$.
Since the gas density effectively involves the density of CO, C, and C$^{+}$, as described above,
the actual H$_{2}$/CO ratio would be even lower than those in Table \ref{tb6}.
It is thus found that the H$_{2}$/CO ratio can be lower than the canonical value (10$^{4}$) for interstellar clouds, 
and the observed line intensities can be reproduced even without H$_{2}$.

A similar analysis is conducted for HD~21997.
Since this source is nearly face-on, the line-of-sight length is approximated by the disk scale height $H$ at 300~au, 
assuming ($H/r$ $\sim$ 0.06):
here we use 20~au in derivation of the averaged number density of CO.
The results are summarized in Table \ref{tb7}.
The optical depths of the CO(2--1) and CO(3--2) lines are much higher in HD~21997 than those found in 49~Ceti.
For this reason, the fitting is not as good as for 49~Ceti, particularly for the blue-shifted component,
where the gas kinetic temperature and the CO column density are not well constrained for the gas density below 10$^{4}$~cm$^{-3}$.
Nevertheless, the gas kinetic temperature is below 12~K, which is similar to the 49~Ceti case.
Again, the H$_{2}$/CO ratio can be lower than the canonical value for interstellar clouds.

We here note two caveats for the above analysis.
First, the optical depth of the CO lines are very high, and hence, the simple approximation by the LVG model \citep{gol74} 
employed here does not perfectly describe the radiation transfer.
In the LVG model, a photon emitted from a certain volume is assumed not to be absorbed in a different part.
This situation may not always be fulfilled for very high optical-depth cases.
This might be the reason for the poor fitting in the HD~21997 case.
More rigorous treatments including dust opacity are left for future study.
For this purpose, we need to resolve the disk structure, and hence, high resolution data of CO and its isotopic species are necessary.
Second, the state-to-state collisional rate coefficients for the CO, C, and C$^{+}$ are not available, and therefore the contribution
of their collisions are not explicitly considered.
Theoretical and experimental studies on the collision rates are awaited.

\subsubsection{The case of electron collision}

Next, we conduct the non-LTE LVG analysis assuming the collision with an electron,
considering that the gas has the secondary origin.
Here, we employ the collisional rate coefficients calculated by using the method described by \cite{dic77}.
As in the H$_{2}$ collision case, we cannot determine the four parameters simultaneously in the analysis, and hence,
we fix the electron density and determine the remaining three parameters.
Because electrons mostly come from photoionization of C (and organic species), the maximum electron density would be around 10$^{4}$~cm$^{-3}$, 
considering the CO number density derived later and the assumption that the C and C$^{+}$ abundances are higher than 
the CO abundance by an order of magnitude as the case of 49~Ceti \citep{rob13, rob14, hig17}.
Since the collision with an electron is more efficient for rotational excitation/de-excitation by two orders of magnitude than that with H$_{2}$ 
(and possibly CO, C, C$^{+}$, and O) \citep{dic77}, the electron abundance relative to CO, C, C$^{+}$, and O needs to be higher than 0.01 
in order for an electron to become a major collision partner.
Hence, we set the minimum electron density of 30~cm$^{-3}$.
The analysis is conducted for several electron densities from 30 to 10$^{4}$~cm$^{-3}$ for 49~Ceti and HD~21997, 
as shown in Tables~\ref{tb8} and \ref{tb9}, respectively.

For 49~Ceti, the column density of CO ranges from 2$\times$10$^{17}$ to 1.2$\times$10$^{18}$~cm$^{-2}$ and 
from 1.8$\times$10$^{17}$ to 9.8$\times$10$^{17}$~cm$^{-2}$ for the blue-shifted and red-shifted components, respectively.
The derived gas kinetic temperature is around 10~K.
These values resemble those found in the H$_{2}$ collision case, except for the electron density of 30~cm$^{-3}$ where the
optical depth of the CO lines is very high due to insufficient excitation.
The beam filling factors also resemble those in the H$_{2}$ collision case.
These similarities seem reasonable because the rotational population is almost in LTE regardless of the collision partner.
For HD~21997, the ranges of the column density and the gas kinetic temperature are also similar to those of the H$_{2}$ collision case.
These results indicate that the CO excitation is also possible without H$_{2}$, if the electron density is higher than 30~cm$^{-3}$.

\subsection{Nature of gas components in debris disks}

Based on the LTE and non-LTE analyses, we confirm that a huge amount of CO gas is associated with the debris disks 
of 49~Ceti and HD~21997, and its gas temperature is quite low ($\sim$10~K).
This result is essentially consistent with the report based on the LTE analysis for HD21997 by \cite{kos13}, but is based on the more
detailed analysis including non-LTE one.
The CO mass\footnote[2]{In order to estimate the CO mass, 
we assume that the disk radius is 3$\arcsec$ and 2$\arcsec$ for 49~Ceti and HD~21997, respectively.} evaluated from the CO column density, 
which is corrected the optical depth effect 
is (6.1 -- 35)$\times$10$^{-2}$$\ME$ and (5.5 -- 85)$\times$10$^{-2}$$\ME$ for 49~Ceti and HD~21997, respectively. 
Our results at the H$_{2}$ density of 1.0$\times$10$^{6}$ to 1.0$\times$10$^{7}$~cm$^{-3}$ and 
electron density of 1.0$\times$10$^{3}$ to 1.0$\times$10$^{4}$~cm$^{-3}$ are roughly consistent with the mass estimate by \cite{kos13} and \cite{moo19}.

The spatial resolution of the present analysis is 6$\arcsec$ for 49~Ceti and 4$\arcsec$ for HD~21997.
Since high angular resolution data of the CO isotopologues are not available, 
we cannot directly derive the radial distribution of the temperature in the disk.
We here discuss the effective temperatures derived above in terms of the temperature distribution of equation \ref{eq1} reported so far.

First, we assume the heating by the central star.
For 49~Ceti, $T_{100}$ of equation 1 is reported to be 40~K for the H$_{2}$/CO ratio of 10$^{4}$ \citep{hug17}.
If the CO molecules distributed in the outer disk make a dominant contribution to the observed emission, the lower gas temperature would be expected.
However, the gas temperature at $R$=300~au is estimated to be 23~K, according to \cite{hug17}, and is still higher than our result.
On the other hand, $T_{100}$ is reported to be 14~K if the H$_{2}$/CO ratio is set to be 1.
If we employ this value, the gas temperature at $R$=300~au is 8~K.
This estimate is close to the gas kinetic temperature derived in this study.
Thus, the low H$_{2}$/CO ratio and the low gas temperature inferred in our analyses (Section 3.3) might be related with each other.
Nevertheless, the temperature structure may not be as simple as equation \ref{eq1}.
We need more detailed analysis based on high spatial resolution observations.

We then consider the effect of the heating by the stellar and interstellar UV radiation on the gas temperature.
Both UV radiation fields play an important role in photodissociation of CO in the outer disk, as shown in 49~Ceti \citep{hig19a}.
The [C\,{\sc i}]/CO intensity ratio decreases with increasing radial distance, has a minimum at 100~au from the central star, and increases again.
However, debris disks contain fewer small grains responsible for photoelectric heating than protoplanetary disks \citep{wil12,mac17}.
For this reason, the gas temperature would not be as high as those in the surface area of typical protoplanetary disks.
Although \cite{bes07} and \cite{zag10} argued that photoelectric heating plays an important role in heating the gas in 
a $\beta$ Pictoris-like debris disk, \cite{kra16} suggest that it does not contribute to the gas heating.
Indeed, the gas temperature in $\beta$ Pictoris is estimated to be as low as 20~K \citep{rob00}.
This is lower than the typical temperature of a diffuse gas in the interstellar clouds illuminated by the interstellar UV radiation ($\sim$ 100~K)
\cite[e.g.,][]{spa96,sno06}.
Thus the low temperature of the gas components seems possible in debris disks even under the exposure to the interstellar UV radiation.

It is well known that CO molecules are depleted onto dust grains at the dust temperature below its desorption temperature ($\sim$ 20~K) \citep{cas99, yam17}.
CO depletion means that CO in the gas phase is frozen out onto dust grains. If the dust temperature is lower than the desorption temperature of CO ($\sim$ 20~K), the CO molecules adsorbed onto dust grain do not come out.
If the dust temperature were low at the outer edge of the disk, CO could be depleted onto dust grains.
The depletion time scale is approximately 0.1~Myr for the gas density of 10$^{4}$~cm$^{-3}$ if the dust-to-gas mass ratio and the dust 
size distribution are the same as those in the interstellar clouds \cite[e.g.,][]{aik12}.
However, even in an outer region, the small dust grains are less abundant in debris disks \citep{paw19} than in the interstellar medium, 
and hence, the CO desorption timescale is extended (see equation 1 of \cite{aik12}).
Hence, the effect of the CO depletion would not be significant.

For the origin of gas, there are mainly two possibilities.
If the gas origin is a primordial case, the H$_{2}$/CO ratio should have been close to the canonical 
be 10$^{4}$ for interstellar clouds or even higher (10$^{5-6}$) as revealed in TW~Hya \citep{ber13, ber18}.
As for the origin of the low H$_{2}$/CO ratio, we first consider the photoevaporation of the primordial gas from a protoplanetary disk.
Photoevaporation \cite[e.g.,][]{nak18} is in general a hydrodynamic escape, where both CO and H$_{2}$ flow out. 
According to the theoretical models, the dispersal timescale was thought to be 3--5~Myr \cite[e.g.,][]{cla01,gor09}.
For forming low H$_{2}$/CO ratio gaseous debris disks,
it is widely considered that Jeans escape \cite[e.g.,][]{vol11} may play a role.
Since the lighter molecules will readily exceed the escape speed, they will preferentially disappear.
Thus, the H$_{2}$ molecules may selectively escape from the debris disk.
This mechanism works better at the lower gas temperature, as found in this study.
However, it is unclear whether it is really possible or not 
in the physical conditions of the debris disks within their ages.


Another possibility is that the gas has a secondary origin, for which several different mechanisms can be considered; e.g., 
the sublimation of dust grains or planetesimals, photo-sputtering of dust grains, collisional vaporization of dust grains, 
and collision of comets or icy planetesimals \cite[e.g.,][]{beu90, gri07, cze07, zuc12, mat17, mar20, kra20}.
\cite{hug17} support this mechanism based on the small scale height of the disk, which is caused by the larger molecular weight due to absence of H$_{2}$.
A recent model explains the observed amount of gas by invoking UV shielding by the C atoms \citep{kra19}.

In order to examine these two possibilities and their relative importance, systematic observations are necessary.
In this regard, we only have a limited amount of observational information.
For example, available archival data for 49~Ceti are of $^{12}$CO, $^{13}$CO, and C$^{18}$O in Band 6 with ACA observations, 
$^{12}$CO in Band 7, and [C\,{\sc i}] in Band 8 with ALMA observations.
For HD~21997, there are $^{12}$CO, $^{13}$CO, and C$^{18}$O in Band 6 and Band 7 with ALMA observations.
In addition to them, it is necessary to perform the high-resolution observations of 
$^{13}$CO(3--2) in Band 7, $^{12}$CO(4--3) and $^{13}$CO(4--3) in Band 8, and $^{12}$CO(7--6) and $^{13}$CO(7--6) 
in Band 10 to derive the distributions of the gas temperature and the CO column densities.
The spatial variation of the CO abundance and its comparison with that of C
will be useful to test various models \cite[e.g.,][]{kra19} for the origin of the gas in debris disks.

Furthermore, there is a possibility that the gas would contain other molecules (CH$_{3}$OH, SO, SiO, H$_{2}$O, etc.). 
These molecules are known as important constituents of icy mantle of dust grains \cite[e.g.,][]{nom09,wal16}, 
and are readily destroyed by UV photodissociation.
If they can be detected in debris disks, the contribution of the secondary gas will be evident. 
Although the outgassing mechanism is not well understood, it is thought to be similar to the process occurring in comets 
(e.g., 67P/Churyumov-Gerasimenko). Thus, volatiles could be ejected from dust grains to the gas component in debris disks. 
If the CH$_{3}$OH abundance is derived, we can discuss it in relation to the observational results reported for various comets \citep{mum11}.
Since a debris disk corresponds to the last phase of planet formation, the result will open a new avenue to study how organic molecules in 
space are incorporated into baby planets.

\section{Summary}

We have analyzed the ALMA archival data of $^{12}$CO and its isotopologues observed toward
the gaseous debris disks, 49~Ceti and HD~21997.
The major results are summarized as follows:

\begin{enumerate}

\item 
We have conducted the LTE and non-LTE analyses to test the excitation of the CO gas
by analyzing the $^{12}$CO, $^{13}$CO, and C$^{18}$O data for 49~Ceti and HD~21997.
For the non-LTE analyses, the two distinct cases of the H$_{2}$ collision and the electron collision are considered.
We have examined a wide range of the gas density, and have determined the beam-averaged column density of CO and the gas temperature.

\item 
The gas kinetic temperature is derived from the non-LTE analyses to be 8--11~K for 49~Ceti for the first time, 
which is significantly lower than the dust temperature.
A similar temperature (8--12~K) is also obtained for HD~21997.
These results would indicate inefficient photoelectric heating due to less abundant small grains.

\item 
The CO column density is derived by the non-LTE analyses to be 
(1.8--5.9)$\times$10$^{17}$~cm$^{-2}$ and (2.6--15)$\times$10$^{17}$~cm$^{-2}$ for 49~Ceti and HD~21997, respectively, 
both for the H$_{2}$ collision and electron collision cases.

\item 
The CO number density averaged over the observation beam is estimated from the
CO column density and the scale of the line of sight.
It is (100--400)~cm$^{-3}$ and (860--5000)~cm$^{-3}$ for 49~Ceti and HD~21997, respectively.

\item 
CO molecules can be excited under the environments of no H$_{2}$ or a small number of H$_{2}$ molecules or even without H$_{2}$.
The wide range of  H$_{2}$ or electron density can account for the CO excitation.

\end{enumerate}

\bigskip
\acknowledgments
We thank the referee for the thoughtful and constructive comments.
This paper makes use of the following ALMA data:
ADS/JAO.ALMA$\#$2011.0.00780.S, 
ADS/JAO.ALMA$\#$2012.1.00195.S, 
and ADS/JAO.ALMA$\#$2016.2.00200.S.
ALMA is a partnership of ESO (representing its member states), NSF (USA) and NINS (Japan), together with NRC (Canada), 
NSC and ASIAA (Taiwan), and KASI (Republic of Korea), in cooperation with the Republic of Chile. 
The Joint ALMA Observatory is operated by ESO, AUI/NRAO and NAOJ.
This study is supported by KAKENHI (18K03713, 18H05222, and 19H05090).
A.M. acknowledges the support of the Hungarian National Research, Development and Innovation Office NKFIH Grant KH-130526. 
Data analysis was carried out on the Multi-wavelength Data Analysis System operated by the Astronomy Data Center (ADC), National Astronomical Observatory of Japan.

\facility{ALMA}
\software{CASA 4.7.2, CASA 5.3.0}

\begin{deluxetable}{l l l l l l l c c c c c c}
\tabletypesize{\scriptsize}
\tablecaption{Data sets}
\tablewidth{0pt}
\tablehead{
\colhead{Object} & \colhead{Line} & \colhead{ALMA ID} &  \colhead{Synthesized Beam} &  \colhead{Array} & \colhead{Reference}}
\startdata
49~Ceti & $^{12}$CO($J$=3--2) & 2012.1.00195.S & 6\farcs4 $\times$ 4\farcs9 (P.A.= $-$ 80.1$^{\circ}$) &12~m & \cite{hug17} \\
49~Ceti & $^{12}$CO($J$=2--1) & 2016.2.00200.S & 6\farcs2 $\times$ 4\farcs8 (P.A.= $-$ 85.0$^{\circ}$) &ACA & \cite{moo19} \\
49~Ceti & $^{13}$CO($J$=2--1) & 2016.2.00200.S & 6\farcs6 $\times$ 5\farcs2 (P.A.= $-$ 88.9$^{\circ}$) &ACA & \cite{moo19} \\
49~Ceti & C$^{18}$O($J$=2--1) & 2016.2.00200.S & 6\farcs5 $\times$ 5\farcs1 (P.A.= $-$ 86.3$^{\circ}$) &ACA & \cite{moo19} \\
\hline
HD~21997 & $^{12}$CO($J$=3--2) & 2011.0.00780.S & 4\farcs0 $\times$ 4\farcs0 (P.A.= 0.0$^{\circ}$) &12~m & \cite{kos13} \\
HD~21997 & $^{12}$CO($J$=2--1) & 2011.0.00780.S & 4\farcs0 $\times$ 4\farcs0 (P.A.= 0.0$^{\circ}$)  &12~m & \cite{kos13} \\
HD~21997 & $^{13}$CO($J$=3--2) & 2011.0.00780.S & 4\farcs0 $\times$ 4\farcs0 (P.A.= 0.0$^{\circ}$) &12~m & \cite{kos13} \\
HD~21997 & $^{13}$CO($J$=2--1) & 2011.0.00780.S & 4\farcs0 $\times$ 4\farcs0 (P.A.= 0.0$^{\circ}$)  &12~m & \cite{kos13} \\
HD~21997 & C$^{18}$O($J$=2--1) & 2011.0.00780.S & 4\farcs0 $\times$ 4\farcs0 (P.A.= 0.0$^{\circ}$)  &12~m & \cite{kos13} \\
\enddata
\label{tb1}
\tablecomments{List of archival data sets.
For a fair comparison among the observations with different spatial resolution, the synthesized beam size
for the 12~m array data is adjusted to that for the ACA data.}
\end{deluxetable}

\begin{deluxetable}{l l l l l l l l l l c c c c c c c c}
\tabletypesize{\scriptsize}
\tablecaption{Line parameters of $^{12}$CO(3--2), $^{12}$CO(2--1), $^{13}$CO(2--1), C$^{18}$O(2--1) from double Gaussian fits for 49~Ceti}
\tablewidth{0pt}
\tablehead{
\colhead{Line} & \colhead{${T_{\rm{B}}}$} & \colhead{$dv$} & \colhead{$V_{\rm{LSR}}$} & \colhead{$T_{\rm{CAL}}$(LTE)} & \colhead{$T_{\rm{CAL}}$(LVG)} 
& \colhead{$T_{\rm{B}}$ -- $T_{\rm{CAL}}$(LTE)} & \colhead{$T_{\rm{B}}$ -- $T_{\rm{CAL}}$(LVG)} \\
& \colhead{[K]} & \colhead{[km~s$^{-1}$]} & \colhead{[km~s$^{-1}$]}  & \colhead{[K]}  & \colhead{[K]} & \colhead{[K]} & \colhead{[K]}}
\startdata
$^{12}$CO(3--2) & 0.16 (0.01) & 1.5 (0.7) & $-$0.0 (0.7)   & 0.16 (0.01) & 0.16 (0.01) & 0.00 & 0.00   \\
$^{12}$CO(3--2) & 0.18 (0.01) & 1.3 (0.7) & 5.5 (0.7)      & 0.18 (0.02) & 0.18 (0.02) & 0.00 & 0.00   \\
\hline
$^{12}$CO(2--1) & 0.23 (0.01) &  1.9 (0.7) & $-$0.3 (0.7)  & 0.23 (0.01) & 0.23 (0.01) & 0.00 & 0.00 \\
$^{12}$CO(2--1) & 0.27 (0.01) &  1.2 (0.7) & 5.5 (0.7)     & 0.27 (0.02) & 0.27 (0.02) & 0.00 & 0.00   \\
\hline
$^{13}$CO(2--1) & 0.10 (0.01) & 1.7 (0.7) &  $-$0.2 (0.7)  & 0.10 (0.01) & 0.10 (0.01) & 0.00 & 0.00  \\
$^{13}$CO(2--1) & 0.13 (0.01) & 1.3 (0.7) & 5.1 (0.7)      & 0.12 (0.02) & 0.13 (0.02) & 0.01 & 0.00 \\
\hline
C$^{18}$O(2--1) & 0.014 (0.010) & 1.4 (0.7) & $-$0.7 (0.7) & 0.02 (0.01) & 0.02 (0.01) & $-$0.01 & 0.00 \\
C$^{18}$O(2--1) & $<$ 0.008 & $<$ 1.0  & --                & 0.03 (0.02) & 0.02 (0.02) & $-$0.02 & $-$0.01 \\
\hline
\enddata
\label{tb2}
\tablecomments{The numbers in parentheses represent the 1$\sigma$ error.
${T_{\rm{B}}}$ is the brightness temperature, $dv$ is the velocity dispersion.
$T_{\rm{CAL}}$ represents the calculated intensity by the analysis.
For the LVG analysis, we adopted the results at the H$_{2}$ density of 1.0$\times$10$^{7}$~cm$^{-3}$.}
\end{deluxetable}

\begin{deluxetable}{l l l l l l l l l l c c c c c c c c}
\tabletypesize{\scriptsize}
\tablecaption{Line parameters of $^{12}$CO(3--2), $^{12}$CO(2--1), $^{13}$CO(3--2), 
$^{13}$CO(2--1) and C$^{18}$O(2--1) from double Gaussian fits for HD~21997}
\tablewidth{0pt}
\tablehead{
\colhead{Line} & \colhead{${T_{\rm{B}}}$} & \colhead{$dv$} & \colhead{$V_{\rm{LSR}}$} & \colhead{$T_{\rm{CAL}}$(LTE)} 
& \colhead{$T_{\rm{CAL}}$(LVG)}  & \colhead{$T_{\rm{B}}$ -- $T_{\rm{CAL}}$(LTE)} & \colhead{$T_{\rm{B}}$ -- $T_{\rm{CAL}}$(LVG)} \\
& \colhead{[K]} & \colhead{[km~s$^{-1}$]} & \colhead{[km~s$^{-1}$]}& \colhead{[K]} & \colhead{[K]} & \colhead{[K]} 
& \colhead{[K]}}
\startdata
$^{12}$CO(3--2) & 0.14 (0.01) & 1.1 (0.1) &  $-$0.3 (0.1) & 0.17 (0.05) & 0.18 (0.07) & $-$0.03 & $-$0.04 & \\
$^{12}$CO(3--2) & 0.14 (0.01) & 1.0 (0.1) &  3.0 (0.1)    & 0.17 (0.04) & 0.18 (0.07) & $-$0.03 & $-$0.04 & \\
\hline
$^{12}$CO(2--1) & 0.29 (0.01) & 1.0 (0.1) &  $-$0.4 (0.1) & 0.26 (0.05) & 0.26 (0.07) & 0.03　& 0.03 & \\
$^{12}$CO(2--1) & 0.28 (0.01) & 1.0 (0.1) &  2.9 (0.1)    & 0.26 (0.04) & 0.25 (0.07) & 0.03  & 0.03  &  \\
\hline
$^{13}$CO(3--2) & 0.10 (0.01) & 0.9 (0.1) &  $-$0.5 (0.1) & 0.08 (0.05) & 0.08 (0.07) & 0.02 & 0.02 & \\
$^{13}$CO(3--2) & 0.10 (0.01) & 1.1 (0.1) &  2.9 (0.1)    & 0.08 (0.04) & 0.08 (0.07) & 0.02 & 0.02  &  \\
\hline
$^{13}$CO(2--1) & 0.15 (0.01) & 0.8 (0.1) &  $-$0.5 (0.1) & 0.18 (0.05) & 0.18 (0.07) & $-$0.03 & $-$0.03  & \\
$^{13}$CO(2--1) & 0.13 (0.01) & 1.1 (0.1) &  2.9 (0.1)    & 0.15 (0.04) & 0.15 (0.07) & $-$0.02 & $-$0.02 &  \\
\hline
C$^{18}$O(2--1) & 0.07 (0.01) & 0.8 (0.1) &  $-$0.4 (0.1) & 0.04 (0.05) & 0.04 (0.07) & 0.03 &  0.03 & \\
C$^{18}$O(2--1) & 0.06 (0.01) & 1.0 (0.1) &  2.9 (0.1)    & 0.03 (0.04) & 0.03 (0.07) & 0.03 &  0.03 & \\
\hline
\enddata
\label{tb3}
\tablecomments{The numbers in parentheses represent the 1$\sigma$ error.
${T_{\rm{B}}}$ is the brightness temperature, $dv$ is the velocity dispersion.
$T_{\rm{CAL}}$ represents the calculated intensity by the analysis.
For the LVG analysis, we adopted the results at the H$_{2}$ density of 1.0$\times$10$^{7}$~cm$^{-3}$.}
\end{deluxetable}

\begin{deluxetable}{l l l l l l l l l l l l c c c c c c c c c c}
\tabletypesize{\scriptsize}
\tablecaption{Physical Parameters (LTE analysis) : 49~Ceti}
\tablewidth{0pt}
\tablehead{
\colhead{${N_{\rm{CO}}}$} & \colhead{${T_{\rm{rot}}}$} & \colhead{$f_{d}$} & \colhead{$\tau_{^{12}\rm{CO}(2-1)}$} 
& \colhead{$\tau_{^{12}\rm{CO}(3-2)}$} & \colhead{$\tau_{^{13}\rm{CO}(2-1)}$} & \colhead{$\tau_{\rm{C^{18}O}(2-1)}$}  \\
\colhead{[cm$^{-2}$]} & \colhead{[K]} & & & & &}
\startdata
[$V_{\rm{LSR}}$=0.0~km~s$^{-1}$] \\
2.2 (1.1) $\times$10$^{17}$ & 9 (1) & 0.06 (0.01) & 38 & 24 & 0.53 & 0.09 \\
\hline
[$V_{\rm{LSR}}$=5.5~km~s$^{-1}$] \\
1.6 (0.5) $\times$10$^{17}$ & 8 (2) & 0.08 (0.04) & 46 & 18 & 0.53 & 0.10 \\
\enddata
\label{tb4}
\tablecomments
{The numbers in parentheses represent the 1$\sigma$ error.
${N_{\rm{CO}}}$ is the CO column density, ${T_{\rm{rot}}}$ is the rotational temperature, $f_{d}$ is the filling factor.
The maximum correlation coefficient is 0.99 between ${T_{\rm{rot}}}$ and $f_{d}$. }
\end{deluxetable}

\begin{deluxetable}{l l l l l l l l l l l l c c c c c c c c c c}
\tabletypesize{\scriptsize}
\tablecaption{Physical Parameters (LTE analysis) : HD~21997}
\tablewidth{0pt}
\tablehead{
\colhead{${N_{\rm{CO}}}$} & \colhead{${T_{\rm{rot}}}$} & \colhead{$f_{d}$} & \colhead{$\tau_{^{12}\rm{CO}(2-1)}$} & 
\colhead{$\tau_{^{12}\rm{CO}(3-2)}$} & \colhead{$\tau_{^{13}\rm{CO}(2-1)}$} & \colhead{$\tau_{^{13}\rm{CO}(3-2)}$} & 
\colhead{$\tau_{\rm{C^{18}O}(2-1)}$}  \\
\colhead{[cm$^{-3}$]} &  \colhead{[K]}  & & & & & & &}
\startdata
[$V_{\rm{LSR}}$=$-$0.3~km~s$^{-1}$] \\
1.9 (0.9) $\times$10$^{17}$ & 8 (3)  & 0.08 (0.06) & 66 & 26 & 1.1 & 0.54 & 0.14 &  \\
\hline
[$V_{\rm{LSR}}$=3.0~km~s$^{-1}$] \\
2.2 (1.0) $\times$10$^{17}$ & 8 (3) & 0.07 (0.05)  & 76 & 36 & 0.87 & 0.56 & 0.13 &  \\
\enddata
\label{tb5}
\tablecomments{The numbers in parentheses represent the 1$\sigma$ error.
${N_{\rm{CO}}}$ is the CO column density, ${T_{\rm{rot}}}$ is the rotational temperature, $f_{d}$ is the filling factor.
The maximum correlation coefficient is 0.99 between ${T_{\rm{rot}}}$ and $f_{d}$.}
\end{deluxetable}

\begin{deluxetable}{l l l l l l l l l l l l l c c c c c c c c c c c}
\tabletypesize{\tiny}
\rotate
\tablecaption{Derived Parameters (Case of H$_{2}$ collision\tablenotemark{a}, LVG analysis) : 49~Ceti}
\tablewidth{0pt}
\tablehead{
\colhead{$n$(H$_{2}$)} & \colhead{${N_{\rm{CO}}}$} & \colhead{${T_{\rm{kin}}}$} & \colhead{$f_{d}$} & 
\colhead{$\tau_{^{12}\rm{CO}(2-1)}$} & \colhead{$\tau_{^{12}\rm{CO}(3-2)}$} & \colhead{$\tau_{^{13}\rm{CO}(2-1)}$} & 
\colhead{$\tau_{\rm{C^{18}O}(2-1)}$} 
& \colhead{$n$(CO)} & \colhead{$n$(H$_{2}$)/$n$(CO)} & \colhead{$M$(CO)} \\
\colhead{[cm$^{-3}$]} & \colhead{[cm$^{-2}$]} & \colhead{[K]} & & & & & & \colhead{[cm$^{-3}$]} & & \colhead{[$\ME$]}}
\startdata
[$V_{\rm{LSR}}$=0.0~km~s$^{-1}$] \\
3.0$\times$10$^{3}$ & 5.9 (1.0) $\times$10$^{17}$ & 11 (1) & 0.04 (0.01) & 110 & 72 & 2.0 & 0.31 & 3.9 (0.7) $\times$10$^{2}$ & 7.6 (1.3) & 9.4 (1.6) $\times$10$^{-2}$ \\
5.0$\times$10$^{3}$ & 4.2 (0.5) $\times$10$^{17}$ & 10 (1) & 0.04 (0.01) & 81 & 50 & 1.4 & 0.20  & 2.8 (0.3) $\times$10$^{2}$ & 18 (2)    & 6.7 (0.8) $\times$10$^{-2}$ \\
1.0$\times$10$^{4}$ & 3.0 (0.3) $\times$10$^{17}$ & 10 (1) & 0.05 (0.01) & 60 & 35 & 0.92 & 0.13 & 2.0 (0.2) $\times$10$^{2}$ & 50 (5)    & 4.8 (0.5) $\times$10$^{-2}$ \\
1.0$\times$10$^{5}$ & 2.1 (0.2) $\times$10$^{17}$ & 9 (1) & 0.05 (0.01) & 45 & 23 & 0.57 & 0.080 & 1.4 (0.1) $\times$10$^{2}$ & 7.1 (0.5) $\times$10$^{2}$ & 3.4 (0.3) $\times$10$^{-2}$ \\
1.0$\times$10$^{6}$ & 2.0 (0.2) $\times$10$^{17}$ & 9 (1) & 0.06 (0.01) & 43 & 22 & 0.55 & 0.076 & 1.3 (0.1) $\times$10$^{2}$ & 7.5 (0.6) $\times$10$^{3}$ & 3.2 (0.3) $\times$10$^{-2}$ \\
1.0$\times$10$^{7}$ & 2.0 (0.2) $\times$10$^{17}$ & 9 (1) & 0.06 (0.01) & 43 & 22 & 0.54 & 0.075 & 1.3 (0.1) $\times$10$^{2}$ & 7.5 (0.6) $\times$10$^{4}$ & 3.2 (0.3) $\times$10$^{-2}$ \\
\hline
[$V_{\rm{LSR}}$=5.5~km~s$^{-1}$] \\
3.0$\times$10$^{3}$ & 5.4 (2.0) $\times$10$^{17}$ & 9 (2) & 0.06 (0.03) & 130 & 72 & 2.5 & 0.33 & 3.6 (1.3) $\times$10$^{2}$  & 8.3 (3.1) & 8.6 (3.2) $\times$10$^{-2}$ \\
5.0$\times$10$^{3}$ & 3.8 (1.2) $\times$10$^{17}$ & 9 (2) & 0.06 (0.03) & 94 & 50 & 1.5 & 0.22 & 2.5 (0.8) $\times$10$^{2}$   & 20 (6)     & 6.1 (1.9) $\times$10$^{-2}$  \\
1.0$\times$10$^{4}$ & 2.7 (0.7) $\times$10$^{17}$ & 9 (2) & 0.07 (0.03) & 69 & 35 & 1.0 & 0.15 & 1.8 (0.5) $\times$10$^{2}$   & 56 (15)    & 4.3 (1.2) $\times$10$^{-2}$  \\
1.0$\times$10$^{5}$ & 1.9 (0.5) $\times$10$^{17}$ & 9 (2) & 0.08 (0.03) & 50 & 22 & 0.64 & 0.089 & 1.3 (0.3) $\times$10$^{2}$ & 7.9 (2.0) $\times$10$^{2}$ & 3.0 (0.8) $\times$10$^{-2}$ \\
1.0$\times$10$^{6}$ & 1.8 (0.5) $\times$10$^{17}$ & 8 (2) & 0.08 (0.04) & 48 & 21 & 0.61 & 0.084 & 1.2 (0.3) $\times$10$^{2}$ & 8.3 (2.0) $\times$10$^{3}$ & 2.9 (0.8) $\times$10$^{-2}$ \\
1.0$\times$10$^{7}$ & 1.8 (0.5) $\times$10$^{17}$ & 8 (2) & 0.08 (0.04) & 48 & 21 & 0.60 & 0.083 & 1.2 (0.3) $\times$10$^{2}$ & 8.3 (2.0) $\times$10$^{4}$ & 2.9 (0.8) $\times$10$^{-2}$ \\
\enddata
\label{tb6}
\tablenotetext{a}{This mimics the primordial gas case.}
\tablecomments{The numbers in parentheses represent the 1$\sigma$ error.
${N_{\rm{CO}}}$ is the CO column density, ${T_{\rm{kin}}}$ is the kinetic temperature, $f_{d}$ is the filling factor.
The maximum correlation coefficient is 0.99 between ${T_{\rm{kin}}}$ and $f_{d}$.}
\end{deluxetable}

\begin{deluxetable}{l l l l l l l l l l l l l c c c c c c c c c c c}
\tabletypesize{\tiny}
\rotate
\tablecaption{Derived Parameters (Case of H$_{2}$ collision\tablenotemark{b}, LVG analysis) : HD~21997}
\tablewidth{0pt}
\tablehead{
\colhead{$n$(H$_{2}$)} & \colhead{${N_{\rm{CO}}}$} & \colhead{${T_{\rm{kin}}}$} & \colhead{$f_{d}$} & \colhead{$\tau_{^{12}\rm{CO}(2-1)}$} & 
\colhead{$\tau_{^{12}\rm{CO}(3-2)}$} & \colhead{$\tau_{^{13}\rm{CO}(2-1)}$} & \colhead{$\tau_{^{13}\rm{CO}(3-2)}$} & \colhead{$\tau_{\rm{C^{18}O}(2-1)}$} &
\colhead{$n$(CO)} & \colhead{$n$(H$_{2}$)/$n$(CO)} & \colhead{$M$(CO)} \\
\colhead{[cm$^{-3}$]} & \colhead{[cm$^{-2}$]} & \colhead{[K]} & & & & & & & \colhead{[cm$^{-3}$]} & & \colhead{[$\ME$]}}
\startdata
[$V_{\rm{LSR}}$=$-$0.3~km~s$^{-1}$] \\
3.0$\times$10$^{3}$ & 1.5 (1.3) $\times$10$^{18}$ & 10 (6)  & 0.05 (0.05) & 430 & 240 & 6.4 & 3.1 & 1.1  & 5.0 (4.3) $\times$10$^{3}$ & 6.0 (5.2) $\times$10$^{-1}$ & 1.6 (1.4) $\times$10$^{-1}$  \\
5.0$\times$10$^{3}$ & 9.5 (6.8) $\times$10$^{17}$ & 10 (6)  & 0.05 (0.05) & 260 & 150 & 3.9 & 2.0 & 0.63 & 3.2 (2.3) $\times$10$^{3}$ & 16 (11)                     & 1.0 (0.7) $\times$10$^{-1}$   \\
1.0$\times$10$^{4}$ & 6.3 (5.0) $\times$10$^{17}$ & 10 (6)  & 0.05 (0.05) & 170 & 100 & 2.4 & 1.4 & 0.37 & 2.1 (1.7) $\times$10$^{3}$ & 48 (38)                     & 6.8 (5.4) $\times$10$^{-2}$  \\
1.0$\times$10$^{5}$ & 3.0 (1.7) $\times$10$^{17}$ & 9 (4) & 0.07 (0.05) & 92 & 45 & 1.2 & 0.60 & 0.16    & 1.0 (5.7) $\times$10$^{3}$ & 1.0 (0.6) $\times$10$^{2}$  & 3.2 (1.8) $\times$10$^{-2}$  \\
1.0$\times$10$^{6}$ & 2.8 (1.5) $\times$10$^{17}$ & 8 (3) & 0.07 (0.05) & 86 & 40 & 1.1 & 0.54 & 0.15    & 9.3 (5.0) $\times$10$^{2}$ & 1.1 (0.6) $\times$10$^{3}$  & 3.0 (1.6) $\times$10$^{-2}$  \\
1.0$\times$10$^{7}$ & 2.8 (1.5) $\times$10$^{17}$ & 8 (3) & 0.07 (0.05) & 86 & 40 & 1.1 & 0.53 & 0.15    & 9.2 (5.0) $\times$10$^{2}$ & 1.1 (0.6) $\times$10$^{4}$  & 3.0 (1.6) $\times$10$^{-2}$  \\
\hline
[$V_{\rm{LSR}}$=3.0~km~s$^{-1}$] \\
3.0$\times$10$^{3}$ & 1.4 (1.3) $\times$10$^{18}$ & 11 (9)  & 0.04 (0.05) & 330 & 230 & 5.3 & 3.1 & 0.93 & 4.6 (4.4) $\times$10$^{3}$ & 6.5 (6.3) $\times$10$^{-1}$ & 1.5 (1.4) $\times$10$^{-1}$  \\
5.0$\times$10$^{3}$ & 9.1 (8.2) $\times$10$^{17}$ & 12 (8)  & 0.03 (0.04) & 200 & 150 & 3.3 & 2.2 & 0.55 & 3.0 (2.7) $\times$10$^{3}$ & 17 (15)    					& 9.8 (8.8) $\times$10$^{-2}$  \\
1.0$\times$10$^{4}$ & 5.5 (4.5) $\times$10$^{17}$ & 12 (8)  & 0.04 (0.04) & 130 & 91 & 1.9 & 1.3 & 0.30  & 1.8 (1.5) $\times$10$^{3}$ & 55 (45)     				& 5.9 (4.8) $\times$10$^{-2}$  \\
1.0$\times$10$^{5}$ & 2.8 (1.7) $\times$10$^{17}$ & 10 (4) & 0.05 (0.04) & 78 & 44 & 1.0 & 0.59 & 0.14   & 9.3 (5.6) $\times$10$^{2}$ & 1.1 (6.4) $\times$10$^{2}$  & 3.0 (1.8) $\times$10$^{-2}$  \\
1.0$\times$10$^{6}$ & 2.6 (1.5) $\times$10$^{17}$ & 9 (4) & 0.06 (0.05) & 74 & 40 & 0.93 & 0.53 & 0.13   & 8.7 (4.9) $\times$10$^{2}$ & 1.2 (6.5) $\times$10$^{3}$  & 2.8 (1.6) $\times$10$^{-2}$  \\
1.0$\times$10$^{7}$ & 2.6 (1.5) $\times$10$^{17}$ & 9 (4) & 0.06 (0.05) & 74 & 40 & 0.92 & 0.53 & 0.13   & 8.6 (4.8) $\times$10$^{2}$ & 1.2 (6.6) $\times$10$^{4}$  & 2.8 (1.6) $\times$10$^{-2}$  \\
\enddata
\label{tb7}
\tablenotetext{b}{This mimics the primordial gas case.}
\tablecomments{The numbers in parentheses represent the 1$\sigma$ error.
${N_{\rm{CO}}}$ is the CO column density, ${T_{\rm{kin}}}$ is the kinetic temperature, $f_{d}$ is the filling factor.
The maximum correlation coefficient is 0.99 between ${T_{\rm{kin}}}$ and $f_{d}$.}
\end{deluxetable}

\begin{deluxetable}{l l l l l l l l l l l l l c c c c c c c c c c c}
\tabletypesize{\tiny}
\rotate
\tablecaption{Derived Parameters (Case of electron collision\tablenotemark{c}, LVG analysis) : 49~Ceti}
\tablewidth{0pt}
\tablehead{
\colhead{$n$(E)} & \colhead{${N_{\rm{CO}}}$} & \colhead{${T_{\rm{kin}}}$} & \colhead{$f_{d}$} & 
\colhead{$\tau_{^{12}\rm{CO}(2-1)}$} & \colhead{$\tau_{^{12}\rm{CO}(3-2)}$} & \colhead{$\tau_{^{13}\rm{CO}(2-1)}$} & 
\colhead{$\tau_{\rm{C^{18}O}(2-1)}$} 
& \colhead{$n$(CO)} & \colhead{$n$(E)/$n$(CO)} & \colhead{$M$(CO)} \\
\colhead{[cm$^{-3}$]} & \colhead{[cm$^{-2}$]} & \colhead{[K]} & & & & & & \colhead{[cm$^{-3}$]} & & \colhead{[$\ME$]} }
\startdata
[$V_{\rm{LSR}}$=0.0~km~s$^{-1}$] \\ 
3.0$\times$10$^{1}$ & 1.2 (0.4) $\times$10$^{18}$ & 11 (3) & 0.04 (0.01) & 210 & 150 & 4.1 &0.66 & 7.9 (3.0) $\times$10$^{2}$ & 3.8 (1.4) $\times$10$^{-2}$ & 1.9 (0.7) $\times$10$^{-1}$ \\
1.0$\times$10$^{2}$ & 4.3 (0.7) $\times$10$^{17}$ & 11 (1) & 0.04 (0.01) & 79 & 54 & 1.4 & 0.20  & 2.9 (0.5) $\times$10$^{2}$ & 3.5 (0.5) $\times$10$^{-1}$ & 7.0 (1.2) $\times$10$^{-2}$ \\
1.0$\times$10$^{3}$ & 2.2 (0.2) $\times$10$^{17}$ & 9 (1) & 0.05 (0.01) & 46 & 25 & 0.61 & 0.085 & 1.5 (0.1) $\times$10$^{2}$ & 6.9 (0.6) & 3.5 (0.3) $\times$10$^{-2}$  \\
1.0$\times$10$^{4}$ & 2.0 (0.2) $\times$10$^{17}$ & 9 (1) & 0.06 (0.01) & 43 & 22 & 0.55 & 0.076 & 1.3 (0.1) $\times$10$^{2}$ & 75 (58)   & 3.2 (0.3) $\times$10$^{-2}$ \\
\hline
[$V_{\rm{LSR}}$=5.5~km~s$^{-1}$] \\
3.0$\times$10$^{1}$ & 9.8 (5.7) $\times$10$^{17}$ & 10 (3) & 0.05 (0.03) & 230 & 140 & 4.0 & 0.63  & 6.5 (3.8) $\times$10$^{2}$ & 4.6 (2.7) $\times$10$^{-2}$ & 1.6 (0.9) $\times$10$^{-1}$ \\
1.0$\times$10$^{2}$ & 3.7 (1.5) $\times$10$^{17}$ & 10 (3) & 0.06 (0.03) & 88 & 51 & 1.4   & 0.21  & 2.5 (1.0) $\times$10$^{2}$ & 4.0 (1.6) $\times$10$^{-1}$ & 6.0 (2.3) $\times$10$^{-2}$ \\
1.0$\times$10$^{3}$ & 2.0 (0.5) $\times$10$^{17}$ & 8 (2)  & 0.07 (0.03) & 51 & 24 & 0.67  & 0.093 & 1.3 (0.4) $\times$10$^{2}$ & 7.7 (2.1) & 3.1 (0.8) $\times$10$^{-2}$ \\
1.0$\times$10$^{4}$ & 1.8 (0.5) $\times$10$^{17}$ & 8 (2)  & 0.08 (0.03) & 48 & 21 & 0.61  & 0.085 & 1.2 (0.3) $\times$10$^{2}$ & 84 (22)   & 2.9 (0.8) $\times$10$^{-2}$ \\
\enddata
\label{tb8}
\tablenotetext{c}{This mimics the secondary gas case.}
\tablecomments{The numbers in parentheses represent the 1$\sigma$ error.
${N_{\rm{CO}}}$ is the CO column density, ${T_{\rm{kin}}}$ is the kinetic temperature, $f_{d}$ is the filling factor.
The maximum correlation coefficient is 0.99 between ${T_{\rm{kin}}}$ and $f_{d}$.}
\end{deluxetable}

\begin{deluxetable}{l l l l l l l l l l l l l c c c c c c c c c c c}
\tabletypesize{\tiny}
\rotate
\tablecaption{Derived Parameters (Case of electron collision\tablenotemark{d}, LVG analysis) : HD~21997}
\tablewidth{0pt}
\tablehead{
\colhead{$n$(E)} & \colhead{${N_{\rm{CO}}}$} & \colhead{${T_{\rm{kin}}}$} & \colhead{$f_{d}$} & \colhead{$\tau_{^{12}\rm{CO}(2-1)}$} & 
\colhead{$\tau_{^{12}\rm{CO}(3-2)}$} & \colhead{$\tau_{^{13}\rm{CO}(2-1)}$} & \colhead{$\tau_{^{13}\rm{CO}(3-2)}$} & \colhead{$\tau_{\rm{C^{18}O}(2-1)}$} &
\colhead{$n$(CO)} & \colhead{$n$(E)/$n$(CO)} & \colhead{$M$(CO)} \\
\colhead{[cm$^{-3}$]} & \colhead{[cm$^{-2}$]} & \colhead{[K]} & & & & & & & \colhead{[cm$^{-3}$]} & & \colhead{[$\ME$]}}
\startdata
[$V_{\rm{LSR}}$=$-$0.3~km~s$^{-1}$] \\
3.0$\times$10$^{1}$ & 3.9 (4.3) $\times$10$^{18}$ & 11 (8)  & 0.04 (0.05) & 930 & 640 & 15  & 9.1  & 2.7     & 1.3 (1.5) $\times$10$^{4}$ & 2.3 (2.6) $\times$10$^{-3}$ & 4.2 (4.7) $\times$10$^{-1}$ \\
1.0$\times$10$^{2}$ & 1.2 (1.2) $\times$10$^{18}$ & 11 (7)  & 0.04 (0.05) & 310 & 210 & 4.7 & 2.9  & 0.77    & 4.1 (4.1) $\times$10$^{3}$ & 2.4 (2.4) $\times$10$^{-2}$ & 1.3 (1.3) $\times$10$^{-1}$ \\
1.0$\times$10$^{3}$ & 3.3 (2.0) $\times$10$^{17}$ & 9 (4)   & 0.06 (0.05) & 98 & 50 & 1.3   & 0.67 & 0.18   & 1.1 (0.7) $\times$10$^{3}$ & 9.1 (5.5) $\times$10$^{-1}$  & 3.5 (2.1) $\times$10$^{-2}$ \\
1.0$\times$10$^{4}$ & 2.8 (1.5) $\times$10$^{17}$ & 8 (3)   & 0.07 (0.05) & 86 & 40 & 1.1   & 0.54 & 0.15  & 9.7 (4.9) $\times$10$^{2}$ & 11 (5.7)                      & 3.0 (1.6) $\times$10$^{-2}$  \\
\hline
[$V_{\rm{LSR}}$=3.0~km~s$^{-1}$] \\
3.0$\times$10$^{1}$ & 4.0 (5.9) $\times$10$^{18}$ & 14 (13)  & 0.03 (0.04) & 820 & 730 & 15 & 11 & 2.9    & 1.3 (2.0) $\times$10$^{4}$ & 2.2 (3.3) $\times$10$^{-3}$ & 4.3 (6.3) $\times$10$^{-1}$ \\
1.0$\times$10$^{2}$ & 1.2 (1.5) $\times$10$^{18}$ & 13 (11)  & 0.03 (0.04) & 260 & 210 & 4.3 & 3.2 & 0.74 & 3.9 (5.0) $\times$10$^{3}$ & 2.6 (3.3) $\times$10$^{-2}$ & 1.2 (1.6) $\times$10$^{-1}$ \\
1.0$\times$10$^{3}$ & 2.8 (1.6) $\times$10$^{17}$ & 10 (5)   & 0.05 (0.05) & 83 & 49 & 1.1 & 0.67 & 0.15  & 9.4 (6.1) $\times$10$^{2}$ & 1.1 (0.7)					 & 3.0 (2.0) $\times$10$^{-2}$ \\
1.0$\times$10$^{4}$ & 2.4 (1.3) $\times$10$^{17}$ & 9 (4)    & 0.06 (0.05) & 74 & 40 & 0.93 & 0.53 & 0.13 & 7.9 (4.4) $\times$10$^{2}$ & 13  (7.1) 					 & 2.5 (1.4) $\times$10$^{-2}$ \\
\enddata
\label{tb9}
\tablenotetext{d}{This mimics the secondary gas case.}
\tablecomments{The numbers in parentheses represent the 1$\sigma$ error.
${N_{\rm{CO}}}$ is the CO column density, ${T_{\rm{kin}}}$ is the kinetic temperature, $f_{d}$ is the filling factor.
The maximum correlation coefficient is 0.99 between ${T_{\rm{kin}}}$ and $f_{d}$.}
\end{deluxetable}

\clearpage

\end{document}